\documentclass{amsart}

\usepackage{latexsym}
\usepackage{amssymb}
\newtheorem{thm}{Theorem}
\newtheorem{prop}{Proposition}
\newtheorem{lemma}{Lemma}
\newtheorem{cor}{Corollary}
\newtheorem{definition}{Definition}

\renewcommand{\Sp}{\ensuremath{\Sigma_{+}}}
\newcommand{\Sm}{\ensuremath{\Sigma_{-}}}
\newcommand{\No}{\ensuremath{N_{1}}}
\newcommand{\Nt}{\ensuremath{N_{2}}}
\newcommand{\Nth}{\ensuremath{N_{3}}}

\begin{document}
\title{Curvature blow up in Bianchi VIII and IX vacuum spacetimes}   
\author{Hans  Ringstr\"{o}m}
\address{Department of Mathematics\\ 
Royal Institute of Technology\\
S-100 44 Stockholm\\ 
Sweden}

\begin{abstract}
The maximal globally hyperbolic development of non-Taub-NUT Bianchi
IX vacuum initial data and of non-NUT Bianchi VIII vacuum initial
data is $C^{2}$-inextendible. Furthermore, a curvature invariant is
unbounded in the incomplete directions of inextendible causal 
geodesics. 
\end{abstract}
\maketitle

\section{Introduction}

According to a conjecture by Belinskii, Khalatnikov and Lifshitz,
the Bianchi IX spacetimes are a good model for the local behaviour
of generic gravitational collapse, see Berger \textit{et al} (1998) and 
references therein. Although they have been analyzed numerically,
few statements concerning these spacetimes have been proven.

There are several formulations of Einstein's equations for Bianchi
IX models. One is the Hamiltonian description due to Misner and
another is the formulation by  Wainwright and Hsu. A brief 
explanation of the different variables can be found 
in Hobill \textit{et al} (1994). We will use the formulation developed
by Wainwright and Hsu (1989). In section \ref{S:whsu} we explain 
how these variables are obtained. 

In order to formulate the main result of this paper we need to define
what is meant by the different types of initial data. Let $G$ be a 
$3$-dimensional Lie group,
$e_{i}$, $i=1,2,3$ be a basis of the Lie algebra with structure
constants determined by  $[e_{i},e_{j}]=\gamma_{ij}^{k}e_{k}$. If
$\gamma_{ik}^{k}=0$, then the Lie algebra and Lie group are said to 
be of class A and  
\begin{equation}\label{eq:sconstants}
\gamma_{ij}^{k}=\epsilon_{ijm}n^{km}
\end{equation}
where the symmetric matrix $n^{ij}$ is given by
\begin{equation}\label{eq:ndef}
n^{ij}=\frac{1}{2}\gamma^{(i}_{kl}\epsilon_{}^{j)kl}.
\end{equation}
\begin{definition}\label{def:data}
Class A vacuum initial data for Einstein's equations consist of
the following. A  Lie group $G$ of class A, a left invariant metric
$g$ on $G$ and a left invariant symmetric covariant two-tensor $k$ on
$G$ satisfying
\begin{equation}\label{eq:con1}
R_{g}-k_{ij}k^{ij}+(\mathrm{tr}_{g} k)^2=0
\end{equation}
and
\begin{equation}\label{eq:con2}
\nabla_{i}\mathrm{tr}_{g} k-\nabla^{j}k_{ij}=0
\end{equation}
where $\nabla$ is the Levi-Civita connection of $g$ and $R_{g}$ is
the corresponding scalar curvature, indices are raised and lowered
by $g$.
\end{definition}
We can choose a left invariant orthonormal basis $\{ e_{i}\}$ with 
respect to $g$ so that the corresponding matrix $n^{ij}$ defined in 
(\ref{eq:ndef}) is diagonal with diagonal elements $n_{1}$, $n_{2}$ 
and $n_{3}$. By an appropriate choice of orthonormal basis 
$n_{1}, n_{2}, n_{3}$ can be assumed to belong to one and only
one of the types given in table \ref{table:bianchiA}. We 
assign a Bianchi type to the initial data accordingly. 
This division constitutes a classification of the class A Lie
algebras. We refer to Lemma \ref{lemma:liealg} for a  proof of 
these statements.

Let $k_{ij}=k(e_{i},e_{j})$. Then the matrices $n^{ij}$
and $k_{ij}$ commute according to (\ref{eq:con2}) so that we may
assume $k_{ij}$ to be diagonal with diagonal elements
$k_{1}$, $k_{2}$ and $k_{3}$, cf. (\ref{eq:transform}). 
\begin{definition}\label{def:generic}
Class A vacuum initial data, except type I and VI$I_{0}$
data with $\mathrm{tr}_{g} k=0$, satisfying $k_{2}=k_{3}$ and
$n_{2}=n_{3}$ or one of the permuted conditions are said to be
non-generic. In the Bianchi IX and VIII cases we call such data  
Taub-NUT and NUT vacuum initial data respectively.
\end{definition}
We will justify this definition in the remark following Lemma
\ref{lemma:development}. Observe that the condition is 
independent of the choice of orthonormal basis diagonalizing 
$n$ and $k$, cf. (\ref{eq:transform}). What 
is meant by inextendibility is explained in the following
\begin{definition}
Consider a connected Lorentz manifold $(M,g)$.
If there is a connected $C^{2}$ Lorentz manifold $(\hat{M},\hat{g})$
of the same dimension and a map $i:M\rightarrow \hat{M}$, with
$i(M)\neq \hat{M}$, which is an isometry onto its image, then  
$(M,g)$ is said to be $C^{2}$-extendible and $(\hat{M},\hat{g})$ is
called a $C^{2}$-extension of $(M,g)$. A Lorentz manifold which is not
$C^{2}$-extendible is said to be $C^{2}$-inextendible. 
\end{definition}
\textit{Remark}. There is an analogous definition of smooth
extensions. Unless otherwise mentioned manifolds are assumed to 
be smooth and maps between manifolds are assumed to be as regular as
possible.

We are now in a position to state the main theorem.
\begin{thm}\label{thm:main}
Consider the maximal globally hyperbolic development of non-Taub-NUT
Bianchi IX vacuum initial data or of non-NUT Bianchi VIII vacuum
initial data. It is $C^{2}$-inextendible and the Kretschmann scalar 
\[
\kappa=R_{\alpha\beta\gamma\delta}R^{\alpha\beta\gamma\delta}
\]
is unbounded in the incomplete directions of inextendible causal 
geodesics.
\end{thm}
\textit{Remark}. One can time orient the Bianchi VIII development 
mentioned in Theorem \ref{thm:main} so that all causal geodesics 
are future complete and past incomplete. 
In the Bianchi IX development mentioned, all causal geodesics are 
future and past incomplete. A proof, not the first, is 
given in Lemma \ref{lemma:development}.

\begin{table}\label{table:bianchiA}
\caption{Bianchi class A.}
\begin{tabular}{@{}lccc}
Type & $n_{1}$ & $n_{2}$ & $n_{3}$ \\
I                   & 0 & 0 & 0 \\
II                  & + & 0 & 0 \\
V$\mathrm{I}_{0}$   & 0 & + & $-$ \\
VI$\mathrm{I}_{0}$  & 0 & + & + \\
VIII                & $-$ & + & + \\
IX                  & + & + & + \\
\end{tabular}
\end{table}

We also obtain qualitative information concerning the asymptotic 
behaviour of solutions in the causally geodesically incomplete
directions. A solution which is not NUT nor Taub-NUT must oscillate 
indefinitely in the variables of Wainwright and Hsu as one approaches 
a singularity.

In his article concerning the global dynamics of the Mixmaster
model, Rendall (1997) proved, among other things, the corresponding 
theorem for Bianchi 
type I (with $\mathrm{tr}_{g}k\neq 0$), II, V$\mathrm{I}_{0}$ and 
VI$\mathrm{I}_{0}$ ($\mathrm{tr}_{g}k\neq 0$) vacuum initial data, 
the exceptional cases being the non-generic data. Bianchi type I 
and VI$\mathrm{I}_{0}$ vacuum
initial data satisfying $\mathrm{tr}_{g}k=0$ yield  geodesically
complete, and consequently inextendible,  maximal globally hyperbolic 
developments, see Lemma \ref{lemma:development}. Thus we have
\begin{cor}
Strong Cosmic Censorship holds for Bianchi class A vacuum initial
data. That is, the maximal globally hyperbolic development of generic 
vacuum initial data of class A is $C^{2}$-inextendible.
\end{cor}
For a related paper see Chru\'{s}ciel and Rendall (1995). 
They isolate the locally homogeneous vacuum initial data on compact 
manifolds whose maximal globally hyperbolic developments have 
smooth extensions. In the Bianchi IX case, they prove that the
homogeneous vacuum initial data that  lead to an extendible maximal 
globally hyperbolic development are the Taub-NUT initial data. 
They also show, in the compact, locally homogeneous case, that
extendibility implies additional symmetry of the solution in
the sense that the local Killing algebra has dimension at least four.
Thus they are able to conclude that extendibility is non-generic.

Our arguments can be extended to a more general situation. Consider
initial data as in Definition \ref{def:data} with $G$ simply
connected. If $\Gamma$ is a subgroup of $G$ that acts freely and 
properly discontinuously on $G$ on the left, then we get well
defined initial data on the quotient $G/\Gamma$ and our results
apply to them. However, one does not necessarily obtain all 
quotients of $G$ that admit locally homogeneous data in this way. 
One can for instance take quotients of $\mathbb{R}^{3}$
by a subgroup of the isometry group (of the standard flat metric) which
is not a subgroup of the group of translations. In certain situations
the process of taking the quotient yields restrictions on the possible
initial data. These restrictions have to be analyzed for one to be able
to answer questions concerning cosmic censorship. We refer to 
Chru\'{s}ciel and Rendall (1995) for further elaboration on this
point.

Section \ref{S:whsu} contains a sketch of a derivation of the 
evolution equations. The first form of the 
equations presented there is due to Ellis and MacCallum (1969). We 
then reformulate them as in the article by Wainwright and Hsu (1989)
and state some curvature expressions we will need later. The 
remaining sections contain the proof.

\section{The equations of Wainwright and Hsu}\label{S:whsu}

In this section we consider a special class of spatially homogeneous
four dimensional vacuum spacetimes of the form
\begin{equation}\label{eq:structure}
(I\times G,-d t^2+\chi_{ij}(t)\xi^{i}\otimes \xi^{j})
\end{equation}
where $I$ is an open interval, $G$ is a Lie group of class A, 
$\chi_{ij}$ is a smooth positive definite matrix and the $\xi^{i}$ are 
the duals of a left invariant basis on $G$.
Below, Latin indices will be raised and lowered by $\delta_{ij}$.

\begin{lemma}\label{lemma:liealg}
Table \ref{table:bianchiA} constitutes a classification of the class
A Lie algebras. Consider an arbitrary basis $\{e_{i}\}$ of the Lie 
algebra. Then by applying an orthogonal matrix to it, we can construct 
a basis $\{e_{i}'\}$ such that the corresponding $n'$ defined by
(\ref{eq:ndef}) has diagonal elements of one of the types given in
table \ref{table:bianchiA}.
\end{lemma}
\textit{Proof}. Let $e_{i}$ be a 
basis for the Lie algebra and $n$ be defined as in (\ref{eq:ndef}).
If we change the basis according to 
$e_{i}'=(A^{-1})_{i}^{\ j}e_{j}$ then $n$ transforms to
\begin{equation}\label{eq:transform}
n'=(\det A)^{-1}A^{t}nA
\end{equation}
Since $n$ is symmetric we assume from here on that the basis is such 
that it is diagonal. The matrix $A=\mathrm{diag}(1\ 1\ -1)$ changes 
the sign of $n$. A suitable orthogonal matrix performs even 
permutations of the diagonal. The number of non-zero elements on the 
diagonal is invariant under transformations (\ref{eq:transform})
taking one diagonal matrix to another. If $A=(a_{ij})$ and the 
diagonal matrix $n'$ is constructed as in (\ref{eq:transform}) we have
$n'_{kk}=(\det A)^{-1}\sum_{i=1}^{3}a_{ik}^{2}n_{ii}$ so that if all 
the diagonal elements of $n$ have the same sign the same is true
for $n'$. The statements of the lemma follow. $\Box$

Let $(M,g)=(I\times G,-d t^2+\chi_{ij}(t)\xi^{i}\otimes 
\xi^{j})$ be four dimensional and $G$ be of class A. Let 
$e_{0}=\partial_{t}$ and $e_{i}=a_{i}^{\ j}Z_{j}$, i=1,2,3, be an 
orthonormal basis, where  $a$ is a $C^{\infty}$ matrix valued function
of $t$  and the $Z_{i}$ are the duals of $\xi^{i}$. Let
the matrix $A$ satisfy $e_{0}(A)+AB=0$, $A(0)=\mathrm{Id}$ where 
$B_{ij}=
<\nabla_{e_{0}}e_{i},e_{j}>$ and $\mathrm{Id}$ is the $3\times 3$
identity matrix. Then $A$ is smooth and $SO(3)$ valued 
and if $e_{i}'=A_{i}^{\ j}e_{j}$ then  
$<\nabla_{e_{0}}e_{i}',e_{j}'>=0$.
Assume $<\nabla_{e_{0}}e_{i},e_{j}>=0$. Observe that 
$[Z_{i},e_{0}]=0$. The $e_{i}$ span the tangent space of $G$ and 
$<[e_{0},e_{i}],e_{0}>=0$. Let
\begin{equation}
\theta(X,Y)=<\nabla_{X}e_{0},Y>,
\end{equation}
$\theta_{\alpha \beta}=\theta(e_{\alpha},e_{\beta})$ and
$[e_{\beta},e_{\gamma}]=\gamma_{\beta \gamma}^{\alpha}e_{\alpha}$  
where Greek indices run from $0$ to $3$. The objects $\theta_{\alpha 
\beta}$ and $\gamma_{\beta \gamma}^{\alpha}$ will be viewed as
smooth functions from $I$ to some suitable $\mathbb{R}^{k}$.
We get $\theta_{00}=\theta_{0i}=0$ and $\theta_{\alpha\beta}$
symmetric. We also have $\gamma_{ij}^{0}=\gamma_{0i}^{0}=0$ and
$\gamma_{0j}^{i}=-\theta_{ij}$. We let $n$ be defined as in 
(\ref{eq:ndef}) and
\[
\sigma_{ij}=\theta_{ij}-\frac{1}{3}\theta \delta_{ij}
\]
where we by abuse of notation have written $\mathrm{tr}(\theta)$ as 
$\theta$. 

We compute the Einstein tensor in terms of $n$, $\sigma$ 
and $\theta$. The Jacobi identities for $e_{\alpha}$ yield
\begin{equation}\label{eq:dndt}
e_{0}(n_{ij})-2n_{k(i}^{}\sigma_{j)}^{\ k}+\frac{1}{3}\theta n_{ij}=0.
\end{equation}
The $0i$-components of the Einstein equations are
\begin{equation}\label{eq:commute2}
\sigma_{i}^{\ k}n_{kj}-n_{i}^{\ k}\sigma_{kj}=0.
\end{equation}
Letting  
$b_{ij}=2n_{i}^{\ k}n_{kj}-\mathrm{tr}(n) n_{ij}$ and
$s_{ij}=b_{ij}-\frac{1}{3}\mathrm{tr}(b)\delta_{ij}$
the trace free part of the $ij$ equations are
\begin{equation}\label{eq:dsdt}
e_{0}(\sigma_{ij})+\theta\sigma_{ij}+s_{ij}=0.
\end{equation}
The fact that $R_{00}=0$ yields the Raychaudhuri equation
\begin{equation}\label{eq:raychaudhuri}
e_{0}(\theta)+\theta_{ij}\theta^{ij}=0
\end{equation}
and using this together with the trace of the $ij$-equations yields a
constraint
\begin{equation}\label{eq:constraint1}
\sigma_{ij}\sigma^{ij}+(n_{ij}n^{ij}-\frac{1}{2}\mathrm{tr}(n)^2)=
\frac{2}{3}\theta^2.
\end{equation}
Equations (\ref{eq:dndt})-(\ref{eq:constraint1}) are special cases
of equations given in Ellis and MacCallum (1969). 
At a point $t_{0}$ we may diagonalize $n$ and $\sigma$ simultaneously
since they commute (\ref{eq:commute2}). Rotating $e_{\alpha}$ by the 
corresponding
element of $SO(3)$ yields upon going through the definitions that the
new $n$ and $\sigma$ are diagonal at $t_{0}$. Collect the 
off-diagonal terms of $n$ and $\sigma$ in one vector $v$. By 
(\ref{eq:dndt}) and (\ref{eq:dsdt}) there is a time dependent matrix
$C$ such that $\dot{v}=Cv$ so that $v(t)=0\ \forall t$ since 
$v(t_{0})=0$. Since the rotation was time independent 
$<\nabla_{e_{0}}e_{i},e_{j}>=0$ holds in the new basis. 

In order to prove curvature blow up, we need to relate the 
maximal existence interval for solutions to 
(\ref{eq:dndt})-(\ref{eq:constraint1}) to the maximal globally 
hyperbolic development. One ingredient is the following lemma, the
proof of which can be found in the appendix. 
\begin{lemma}\label{lemma:development}
Let $(G,g,k)$ be class A vacuum initial data. We can then choose
a left invariant orthonormal basis $\{ e_{i}'\}$ with respect to $g$ 
so that the corresponding $n'$ given by (\ref{eq:ndef}) is of one of
the types given in table \ref{table:bianchiA}. We can also assume 
$k_{ij}=k(e_{i}',e_{j}')$ to be diagonal. Then there is a manifold 
as in (\ref{eq:structure}) where 
$I=(t_{-},t_{+})$ is the maximal existence interval for a solution of
(\ref{eq:dndt})-(\ref{eq:constraint1}) with initial data
$n(t_{0})=n'$, $\sigma_{ij}(t_{0})=k_{ij}-\mathrm{tr}(k)\delta_{ij}/3$
and $\theta(t_{0})=\mathrm{tr}(k)$, solving Einstein's vacuum equations,
such that the metric restricted to $M_{t_{0}}=\{t_{0}\}\times G$ is
$g$ and $k$ is the second fundamental form of $M_{t_{0}}$.
The development has the following properties:
\begin{enumerate}
\item Each $M_{v}=\{ v\}\times G$, $v\in I$, is a Cauchy surface.
\item If the initial data is not of type IX, the development can be time
   oriented so that it is future causally geodesically complete and past
   causally geodesically incomplete (unless the data is of type I or
   VI$I_{0}$ with $\mathrm{tr}_{g}k=0$ in which case the development
   is causally geodesically complete). We assume this time orientation
   when we speak of such developments, and always that $\partial_{t}$ is
   future oriented. The development of Bianchi IX initial data is both
   future and past causally geodesically incomplete.
\item The causally geodesically complete developments have $\theta=0$
   and $\sigma_{ij}=0$ for the entire solution. The developments with
   one complete and one incomplete direction have $\theta>0$ for the
   entire solution. Bianchi IX developments have $\theta>0$ in
   $(t_{-},t_{0})$ and $\theta<0$ in $(t_{0},t_{+})$ for some 
   $t_{0}\in (t_{-},t_{+})$.
\item If the Kretschmann scalar $\kappa$ is unbounded as $t\rightarrow
   t_{-}$ then it is unbounded
   along past inextendible causal geodesics. Similarly for
   $t\rightarrow t_{+}$ and future oriented causal geodesics.
\end{enumerate}
\end{lemma}
\textit{Remark}. By inspecting (\ref{eq:dndt}) and (\ref{eq:dsdt})
we see that if $k_{2}=k_{3}$ and $n_{2}'=n_{3}'$ then $\sigma_{2}=
\sigma_{3}$ and $n_{2}=n_{3}$ for the entire solution.
Since such solutions are of NUT and Taub-NUT type for the Bianchi VIII
and IX cases respectively, see Ellis and MacCallum (1969), NUT and
Taub-NUT vacuum initial data yield NUT and Taub-NUT developments 
respectively.

Introduce, as in 
Wainwright and Hsu (1989),
\begin{eqnarray*}
\Sigma_{ij} & = & \sigma_{ij}/\theta \\
N_{ij} & = & n_{ij}/\theta \\
B_{ij} & = & 2N_{i}^{\ k}N_{kj}-N^{k}_{\ k}N_{ij} \\
S_{ij} & = & B_{ij}-\frac{1}{3}B^{k}_{\ k}\delta_{ij}
\end{eqnarray*}
and define a new time coordinate $\tau$, independent of time
orientation, by
\begin{equation}\label{eq:dtdtau}
\frac{d t}{d\tau}=\frac{3}{\theta}.
\end{equation}
For Bianchi IX developments as in Lemma \ref{lemma:development} 
we only consider the part of spacetime where 
$\theta$ is strictly positive or strictly negative. Let
$\Sp=\frac{3}{2}(\Sigma_{22}+\Sigma_{33})$ and 
$\Sm=\sqrt{3}(\Sigma_{22}-\Sigma_{33})/2$. If we let $N_{i}$ be the 
diagonal elements of $N_{ij}$, equations (\ref{eq:dndt}) and 
(\ref{eq:dsdt}) turn into
\begin{eqnarray}
\No' & = & (q-4\Sp)\No  \nonumber \\
\Nt' & = & (q+2\Sp +2\sqrt{3}\Sm)\Nt  \nonumber \\
\Nth' & = & (q+2\Sp -2\sqrt{3}\Sm)\Nth  \label{eq:whsu}\\
\Sp' & = & -(2-q)\Sp-3S_{+}  \nonumber \\
\Sm' & = & -(2-q)\Sm-3S_{-} \nonumber 
\end{eqnarray}
where the prime denotes derivative with respect to $\tau$ and
\begin{eqnarray}
q & = & 2(\Sp^2+\Sm^2) \nonumber \\
S_{+} & = & \frac{1}{2}[(\Nt-\Nth)^2-\No(2\No-\Nt-\Nth)]
\label{eq:whsudef}\\
S_{-} & = & \frac{\sqrt{3}}{2}(\Nth-\Nt)(\No-\Nt-\Nth). \nonumber
\end{eqnarray}
The vacuum constraint (\ref{eq:constraint1}) becomes
\begin{equation}
\Sp^2+\Sm^2+\frac{3}{4}[\No^2+\Nt^2+\Nth^2-2(\No\Nt+\Nt\Nth+\Nth\No)]
=1.
\label{eq:constraint}
\end{equation}
The equations (\ref{eq:whsu})-(\ref{eq:constraint}) have certain
symmetries described in Wainwright and Hsu (1989). By permuting
$\No,\Nt,\Nth$ arbitrarily  we get new solutions if we at the same 
time carry out appropriate
combinations of rotations by integer multiples of $2\pi/3$
and reflections in the $(\Sp,\Sm)$-plane. Below we refer to rotations
by integer multiples of $2\pi/3$ as rotations.
Changing the sign of all the $N_{i}$ at the same time does not change
the equations. Classify points $(\No,\Nt,\Nth,\Sp,\Sm)$ according to
the values of $\No,\Nt,\Nth$ in the same way as in table
\ref{table:bianchiA}. Since the sets 
$N_{i}>0$, $N_{i}<0$ and $N_{i}=0$ are invariant under the flow of 
the equation we may classify solutions to 
(\ref{eq:whsu})-(\ref{eq:constraint}) accordingly. When we speak of 
Bianchi IX solutions we will assume $N_{i}>0$, and for Bianchi VIII 
solutions we will assume two $N_{i}>0$ and one $<0$. 

There are three special points in the $(\Sp,\Sm)$-plane. They are 
$(-1,0)$ and $(1/2,\pm\sqrt{3}/2)$ on the Kasner 
circle $q=2$. They are rotated into one another by applying the 
symmetries and the corresponding type I solutions represent isometric 
flat universes. If the spatial
topology is $\mathbb{R}^{3}$, the metric of these solutions can be
written as
\begin{equation}\label{eq:flatkasner}
d s^{2}=-d t^{2}+\sum_{i=1}^{3}t^{2p_{i}}d x^{i}\otimes d
x^{i}
\end{equation}
on the manifold $M=\mathbb{R}_{+}\times \mathbb{R}^{3}$, where 
$\mathbb{R}^{+}=\{ t\in \mathbb{R} : t>0\}$, with two of the $p_{i}$
zero and one equal to $1$. The rotational symmetries in the 
$\Sp\Sm$-plane correspond to permutational symmetry of the $p_{i}$. 
We see that if for instance $p_{2}=p_{3}=0$ we have a rotational 
symmetry in the $x^{2}x^{3}$-plane at each point of $M$.

The set $\Sm=0$, $\Nt=\Nth$ is invariant under the flow of
(\ref{eq:whsu})-(\ref{eq:constraint}). Applying the symmetries to 
this set we get new invariant sets. These sets correspond to the
non-generic data in Definition \ref{def:generic}. Specifically,
in the Bianchi IX case they correspond to the Taub-NUT solutions.

The Raychaudhuri equation (\ref{eq:raychaudhuri}) takes the form
\begin{equation}\label{eq:raychaudhuri2}
\theta'=-(1+q)\theta.
\end{equation}
According to Lemma \ref{lemma:existence} the existence interval for 
solutions of (\ref{eq:whsu})-(\ref{eq:constraint}) is past infinite.
As $\tau\rightarrow -\infty$,  $\theta(\tau)$ goes 
to infinity exponentially according to (\ref{eq:raychaudhuri2}). 

The normalized Kretschmann scalar is given by
\begin{equation}\label{eq:nks}
\tilde{\kappa}=R_{\alpha\beta\gamma\delta}R^{\alpha\beta\gamma\delta}/
\theta^4.
\end{equation}
We have
\[
\tilde{\kappa}=8(E_{ij}E^{ij}-H_{ij}H^{ij})
\]
where, by Wainwright and Ellis (1997), the normalized $E$ and
$H$ fields are given by
\begin{eqnarray}
E_{ij} & = & \frac{1}{3}\Sigma_{ij}-(\Sigma_{i}^{\ k}\Sigma_{kj}-
\frac{2}{3}\Sigma^2\delta_{ij})+S_{ij} \label{eq:E} \\
H_{ij} & = & -3\Sigma^{k}_{\ (i}N_{j)k}+N^{kl}\Sigma_{kl}\delta_{ij}+
\frac{1}{2}N^{k}_{\ k} \Sigma_{ij} \label{eq:H}
\end{eqnarray}
and $\Sigma^{2}=\Sigma_{ij}\Sigma^{ij}/2$. We may consider 
all matrices involved to be 3-vectors. If we write $E_{ij}$ 
as $E=(E_{1}\  E_{2}\  E_{3})$ and similarly for $H$ we have
$\tilde{\kappa}=8(|E|^2-|H|^2)$. Using the definitions and the 
constraint (\ref{eq:constraint}) we compute
\begin{eqnarray}\label{eq:EH}
H_{1} & = & \No\Sp+\frac{1}{\sqrt{3}}(\Nt-\Nth)\Sm  \\
H_{2} & = & -\frac{1}{2}\Nt(\Sp+\sqrt{3}\Sm)+\frac{1}{2}(\Nth-
\No)(\Sp-\frac{1}{\sqrt{3}}\Sm) \\
E_{2}-E_{3} & = & \frac{2}{3\sqrt{3}}\Sm(1-2\Sp)+(\Nt-\Nth)(\Nt
+\Nth-\No) \\
E_{2}+E_{3} & = & \frac{2}{9}\Sp+\frac{10}{9}\Sp^2+\frac{2}{3}
\Sm^2-\frac{8}{9}+(\Nt-\Nth)^2-\No(\Nt+\Nth).
\end{eqnarray}
The fact that $E$ and $H$ are traceless yields the remaining 
components. We end this section with a technical lemma relating 
the existence intervals for the variables of Wainwright and Hsu
and those of Ellis and MacCallum.
\begin{lemma}\label{lemma:emwh}
If for all non-Taub-NUT Bianchi IX initial data for
(\ref{eq:whsu})-(\ref{eq:constraint}) we can prove the existence
of a sequence $\tau_{k}\rightarrow -\infty$ such that
$\tilde{\kappa}(\tau_{k})$ does not converge to zero, then for 
every development as in Lemma \ref{lemma:development} of non-Taub-NUT
initial data, the Kretschmann scalar is unbounded as $t\rightarrow
t_{-}$ and $t\rightarrow t_{+}$ (here $t_{+}$ and $t_{-}$ are the same
as the ones occurring in Lemma \ref{lemma:development}). Similarly, 
if for all non-NUT Bianchi
VIII initial data of (\ref{eq:whsu})-(\ref{eq:constraint}) we can
prove the existence of a sequence $\tau_{k}\rightarrow -\infty$ such 
that $\tilde{\kappa}(\tau_{k})$ does not converge to zero, then for 
every development as in Lemma \ref{lemma:development} of non-NUT
initial data, the Kretschmann scalar is unbounded as $t\rightarrow
t_{-}$. 
\end{lemma}
\textit{Remark}. That the existence interval for solutions to
(\ref{eq:whsu})-(\ref{eq:constraint}) is past infinite is proven in
Lemma \ref{lemma:existence}.

\section{Elementary properties of solutions}
We begin by giving the past existence intervals for solutions to 
(\ref{eq:whsu})-(\ref{eq:constraint}). Let 
$g\in C^{\infty}(\mathbb{R}^{n},\mathbb{R}^{n})$ and
consider $\dot{x}=g(x)$, $x(0)=x_{0}$. Let $(t_{-},0]$ 
be the maximal past existence interval for the solution.

\begin{lemma}\label{lemma:lpe}
If $t_{-}>-\infty$ there is no sequence $t_{k}\rightarrow t_{-}$,
$t_{k}\in (t_{-},0]$, such that $x(t_{k})$ converges.
\end{lemma}
\textit{Proof}. Assume $t_{k}\rightarrow t_{-}$, $x(t_{k})
\rightarrow x_{-}$. There is an $\epsilon>0$ and a $\delta>0$ such 
that we have a smooth flow 
\[
\Phi:B_{\epsilon}(x_{-})\times (-\delta,
\delta)\rightarrow \mathbb{R}^{n}
\]
where $B_{\epsilon}(x_{-})$ is the open ball with
radius $\epsilon$ and center $x_{-}$. Choose a $k$ such that 
$x(t_{k})\in B_{\epsilon}(x_{-})$ and $|t_{k}-t_{-}|<\delta/2$. 
Define $y(t)=x(t)$ for $t\in(t_{k},0]$ and $y(t_{k}-t)=
\Phi(x(t_{k}),t)$ for $t\in[0,\delta)$. Then $y$ extends $x$ beyond 
the maximal existence interval. $\Box$

\begin{lemma}\label{lemma:existence}
The past existence interval for solutions to 
(\ref{eq:whsu})-(\ref{eq:constraint}) is $(-\infty,0]$.
\end{lemma}
\textit{Proof}. For all types except IX the constraint implies
$\Sp^2+\Sm^2\leq 1$
so that $\No,\Nt,\Nth$ do not grow faster than exponentially,
by (\ref{eq:whsu}). Let $(\tau_{-},\tau_{+})$ be the maximal
existence interval for a specific solution, not of type IX. If
$\tau_{-}>-\infty$ we may for any sequence $\tau_{k}>\tau_{-}$
converging to $\tau_{-}$ extract a subsequence such that
the solution converges, contradicting
Lemma \ref{lemma:lpe}. In other words we may not have 
$\tau_{-}>-\infty$, and in the same way we have $\tau_{+}=\infty$. 
Consider solutions of type IX. Let $\tau\leq 0$. We get
\[
\No(\tau)=\exp \{\int_{0}^{\tau}[q(s)-4\Sp(s)]d s\}\No(0)\leq
e^{-2\tau}\No(0)
\] 
and similarly for $\Nt$ and $\Nth$. Thus, the $N_{i}$
do not grow faster than exponentially as we go backward in time
and by the constraint the same holds for $(\Sp,\Sm)$. The past 
existence interval must be $(-\infty,0]$. $\Box$

We will need a few more lemmas concerning the behaviour of solutions
of type VIII and IX.
\begin{lemma}\label{lemma:B9N}
Consider a solution of type IX. The image $(\Sp,\Sm)((-\infty,0])$ 
is contained in a compact set whose size depends on the initial 
data. Further, if at a point $\Nth\geq\Nt\geq\No$ and $\Nth\geq 2$,
then $\Nt\geq\Nth/10$.
\end{lemma}
\textit{Remark}. The importance of the second part of the
lemma is to be found in the consequence that one of the $N_{i}$
may not go to infinity alone. 

\textit{Proof}. Let us prove the second statement first. Assume
$\Nth\geq 2$ and $\Nth\geq 10\Nt\geq 10\No$, so that $-\Nt\geq
-\Nth/10$ and similarly for $\No$. We get
\[(\No-\Nt)^2+\Nth^2-2\Nth(\No+\Nt)\geq \frac{12}{5}
\]
contradicting the constraint (\ref{eq:constraint}). We are done. 
Since $(\No\Nt\Nth)'=3q\No\Nt\Nth$ the product decreases as time 
decreases, and we have 
\begin{equation}
(\No\Nt\Nth)(\tau)\leq C<\infty \label{eq:productbound}
\end{equation}
for $\tau\leq 0$, where $C$ is a constant. There is a compact set
$K_{1}$ such that if $N_{i}\leq 2$, $i=1,2,3$, then $(\Sp,\Sm)
\in K_{1}$. If one of the $N_{i}$ is greater than 2, 
we may by a permutation assume that $\Nth\geq\Nt\geq\No$ and
by the first part of the lemma conclude that $\Nt\geq\Nth/10$.
Observe that then $\No(\Nt+\Nth)<10C$ so that
\[
\Sp^2+\Sm^2\leq 1+15C
\]
by the constraint.
$\Box$

\begin{lemma}\label{lemma:B8N}
Consider a solution of type VIII. Assume $\Nth\geq\Nt>0>\No$.
Then $\No^2\leq 4/3$. If further $\Nth\geq 4$ then $\Nt\geq \Nth/2$.
\end{lemma}
\textit{Proof}. The constraint says
\begin{equation}\label{eq:shearcon}
\Sp^2+\Sm^2+\frac{3}{4}[\No^2+(\Nt-\Nth)^2-2\No(\Nt+\Nth)]=1.
\end{equation}
Since all terms on the left are non-negative, the first part of 
the lemma is immediate. The second part follows from
\[
\Nth-\Nt\leq\frac{2}{\sqrt{3}}\leq 2\leq \frac{1}{2}\Nth.\ \Box
\]
We end this section with a lemma needed in the sequel. Consider
a positive function $N$ satisfying $N'=hN$ where 
$h(\tau)\rightarrow \alpha$ as $\tau\rightarrow -\infty$

\begin{lemma}\label{lemma:limestimate}
Assume $N$ and $h$ are as above. Then for all $\epsilon>0$ there is
a $T$ such that $\tau\leq T$ implies 
\[
\exp[(\alpha+\epsilon)\tau]\leq N(\tau)\leq \exp[(\alpha-\epsilon)
\tau].
\]
\end{lemma}

\section{Limit characterization of special solutions}

\begin{prop}\label{prop:limchar}
A solution to (\ref{eq:whsu})-(\ref{eq:constraint}) satisfies
\[
\lim_{\tau\rightarrow -\infty}(\Sp (\tau),\Sm (\tau))=(-1,0)
\]
only if it is contained in the invariant set
$\Sm=0$ and $\Nt=\Nth$.
\end{prop}
\textit{Remark}. The implication is in fact an equivalence.
The analogue for $(\Sp,\Sm)\rightarrow (1/2,\pm \sqrt{3}/2)$ is true
due to the symmetries. 

Assume in this section that $(\Sp,\Sm)$ converges to $(-1,0)$.
Consider
\[
f=\frac{4}{3}\Sm^2+(\Nt-\Nth)^2.
\]
Observe that $f$ is either identically zero or always strictly
positive due to the fact that $\Nt=\Nth$, $\Sm=0$ is an invariant
set. A function related 
to this one occurs in Wainwright and
Hsu (1989). We prove an estimate of the form
$f(T)\leq g(\tau,T)$ for some $T$ and $\tau\leq T$ and then
that $g(\tau,T)\rightarrow 0$ as $\tau \rightarrow
-\infty$. Compute
\begin{equation}\label{eq:fpr}
f'=-(2-q)\frac{8}{3}\Sm^2+4\sqrt{3}\No(\Nt-\Nth)\Sm+2(q+2\Sp)
(\Nt-\Nth)^2.
\end{equation}

\begin{lemma}\label{fe}
For all $\epsilon>0$ there is a $T$ such that 
\[ f(T)\leq f(\tau)\exp[\epsilon 
(T-\tau)]
\]
for all $\tau\leq T$.
\end{lemma}
\textit{Proof}. By Lemma \ref{lemma:limestimate} $\No$
goes to zero as  $\tau \rightarrow -\infty$.  The same is true
of $2-q$ and $q+2\Sp$. Thus, for all $\epsilon>0$ there exists a 
$T$ such that $\tau\leq T$ implies $f'\leq \epsilon f$ by
(\ref{eq:fpr}). The lemma follows. $\Box$ 

\begin{lemma}
If there is a sequence $\tau_{k}\rightarrow -\infty$ as 
$k\rightarrow \infty$ such that $\Sp(\tau_{k})\leq -1$, then
$\Sm=0$ and $\Nt=\Nth$.
\end{lemma}
\textit{Proof}. The constraint yields, in the points $\tau_{k}$,
\[
\frac{3}{4}f=\Sm^2+\frac{3}{4}(\Nt-\Nth)^2\leq\frac{3}{2}\No
(\Nt+\Nth).
\]
Applying Lemma \ref{lemma:limestimate} to $\No\Nt$ and $\No\Nth$, 
we have, for $k$ large 
enough, $f(\tau_{k})\leq \exp (2\tau_{k})$. Choose a finite $T$ 
corresponding to $\epsilon=1$ in the previous lemma. We get
$f(T)\leq \exp (\tau_{k}+T)$. Letting $k\rightarrow \infty$ we get 
$f(T)=0$, but then $f$ is identically zero. $\Box$

\begin{lemma}
If there is an $S$ such that $\Sp(\tau)\geq -1$ for all 
$\tau\leq S$ then $\Sm=0$ and $\Nt=\Nth$.
\end{lemma}
\textit{Proof}. Using the constraint to express $2-q$ in the $N_{i}$
we get, if we assume $-1\leq\Sp\leq 0$,
\[
(\Sp+1)'\leq 6\No^2+6|\No(\Nt+\Nth)|.
\]
Let $T$ be such that the right hand side $\leq 2e^{2\tau}$ and 
$-1\leq \Sp(\tau) \leq 0$ for all $\tau\leq T$. Integrate the 
inequality to obtain
\[
0\leq\Sp(\tau)+1\leq e^{2\tau}
\]
for all $\tau\leq T$. But then the constraint yields
\[
\frac{3}{4}f=-\frac{3}{4}\No^2+
\frac{3}{2}\No(\Nt+\Nth)+(1-\Sp)(1+\Sp).
\]
By the above argument we have control over the last term, and as
in the previous lemma we have control over the first two terms.
Thus, there exists an $S'$ such that $\tau\leq S'$ implies
$f(\tau)\leq e^{\tau}$.
We deduce, using $\epsilon=\frac{1}{2}$ in Lemma \ref{fe}, that 
$f$ is identically zero.
$\Box$

\section{Properties of past limit points}\label{section:ppl}

If there is a sequence $\tau_{k}\rightarrow -\infty$ such that
a solution to (\ref{eq:whsu})-(\ref{eq:constraint}) evaluated
at $\tau_{k}$ converges to $x$, then $x$ is said to be an
$\alpha$-limit point of the solution. The $\alpha$-limit set
of a solution is the union of its $\alpha$-limit points. 
Consider a solution of type VIII or IX. Assume it has an 
$\alpha$-limit point $(n_{1},n_{2},n_{3},s_{+},s_{-})$. The 
objective of this section is to prove that if $(s_{+},s_{-})$ is not 
one of the special points it has an $\alpha$-limit point at which 
the normalized Kretschmann scalar is non-zero. The following lemma
will be needed,

\begin{lemma}[Rendall (1997)]\label{lemma:lemma21} Let $U$ be an open 
subset of $\mathbb{R}^{n}$, 
$f\in C^{\infty}(\mathbb{R}^{n},\mathbb{R}^{n})$ 
and let $F:U\rightarrow \mathbb{R}$ be a continuous function such
that $F(x(t))$ is strictly monotone for any solution $x(t)$ of
\begin{equation}\label{eq:lemma21}
x'(t)=f(x(t))
\end{equation}
as long as $x(t)$ is in $U$. Then no solution of (\ref{eq:lemma21})
whose image is contained in $U$ has an $\alpha$-limit point in $U$.
\end{lemma}
As noted in Rendall (1997),

\begin{lemma}\label{lemma:onz}
If a solution of type VIII or IX has an 
$\alpha$-limit point $(n_{1},n_{2},n_{3},s_{+},s_{-})$,
then one of the $n_{i}$ must be zero.
\end{lemma}
\textit{Proof}. We have
\[
|\No\Nt\Nth|'=3q|\No\Nt\Nth|.
\]
If $q=0$ then $|\Sp'|+|\Sm'|>0$ and consequently $|\No\Nt\Nth|$
is strictly decreasing as we go backward in time if it is non-zero. 
The statement follows by Lemma \ref{lemma:lemma21}. $\Box$

\begin{cor}
If a solution of type VIII or IX has an 
$\alpha$-limit point $(n_{1},n_{2},n_{3},s_{+},s_{-})$,
then $s_{+}^{2}+s_{-}^{2}\leq 1$.
\end{cor}
Observe that points of type VI$\mathrm{I}_{0}$ defined by $\Sp=-1$,
$\Sm=0$, $\Nt=\Nth\neq0$ and $\No=0$ are fixed points to 
(\ref{eq:whsu})-(\ref{eq:constraint}). The same is true of the 
rotated points. These points are called flat points of type
VI$\mathrm{I}_{0}$. These solutions correspond to the special points 
in the following sense. The vacuum spacetime given by the manifold
$M=\mathbb{R}_{+}\times\mathbb{R}^{3}$ with a metric as in
(\ref{eq:flatkasner}) where $p_{1}=1$ and $p_{2}=p_{3}=0$
can be viewed as a type VI$\mathrm{I}_{0}$ spacetime by
giving $\mathbb{R}^{3}$ a different group structure. Then the
Wainwright and Hsu variables corresponding to a suitable 
frame (left invariant under the new operation on
$\mathbb{R}^{3}$) will be a flat point of type VI$\mathrm{I}_{0}$.

\begin{thm}[Rendall (1997)]\label{theorem:rendall}
The $\alpha$-limit set of a solution of 
(\ref{eq:whsu})-(\ref{eq:constraint}) of type I, II, V$I_{0}$ or 
VI$I_{0}$ is a single point of type I or a flat point of type 
VI$I_{0}$, the latter only being possible if the solution is time 
independent.
\end{thm}

\begin{lemma}\label{lemma:knz}
Consider a solution to (\ref{eq:whsu})-(\ref{eq:constraint}) of 
type VIII or IX. If it has an $\alpha$-limit point
$(n_{1},n_{2},n_{3},s_{+},s_{-})$ such that $(s_{+},s_{-})$ is not
special, it has at least two $\alpha$-limit points on the Kasner 
circle at least one of which is non-flat. In consequence there is
an $\alpha$-limit point such that the normalized Kretschmann scalar 
evaluated at it is non-zero.
\end{lemma}
\textit{Proof}. The solution of (\ref{eq:whsu})-(\ref{eq:constraint})
with initial value $(n_{1},n_{2},n_{3},s_{+},s_{-})$ consists of 
$\alpha$-limit points to the original solution. Let us denote it
$(\No,\Nt,\Nth,\Sp,\Sm)$. The different
solutions will be referred to as the solution and the original 
solution. Let us first prove that there is a non-flat $\alpha$-limit
point on the Kasner circle.
Since $(s_{+},s_{-})$ is not special we know that the 
solution is not a flat point of type VI$\mathrm{I}_{0}$ and we can 
consequently assume that the $\alpha$-limit set of the solution is a 
single point of type I by Theorem \ref{theorem:rendall}, 
say $(0,0,0,\sigma_{+},\sigma_{-})$. If
$(\sigma_{+},\sigma_{-})$ is not special we are done so
assume it is $(-1,0)$. By arguments given in the proof of
Theorem \ref{theorem:rendall},
$\No,\Nt,\Nth$  have to belong to a compact set as we go backward 
in time. If $(\Sp,\Sm)$ does not converge to $(-1,0)$ 
we may thus construct an $\alpha$-limit point to the solution
different from the only one that exists. By Proposition 
\ref{prop:limchar} we conclude that $\Sm=0$ and $\Nt=\Nth$. If 
$\No=0$ we get $\Sp=\pm 1$ by the constraint so that $\Sp=-1$ 
contradicting the fact that $(s_{+},s_{-})$ is not special. Thus 
$\No\neq 0$ and $\Nt=\Nth=0$. Applying the flow to this point we 
conclude that 
$\lim_{\tau\rightarrow \infty}(\Sp(\tau),\Sm(\tau))=(1,0)$, again
the details are to be found in the proof of Theorem 
\ref{theorem:rendall}.
Since the $\alpha$-limit set is closed we conclude that the
original solution has a non-flat $\alpha$-limit point on the Kasner
circle.

We now prove that there are at least two $\alpha$-limit points on
the Kasner circle. Given the non-flat limit point $(\sigma_{+},
\sigma_{-})$, there is a neighbourhood $B_{\epsilon}(\sigma_{+},
\sigma_{-})$ of it in the $\Sp\Sm$-plane such that two $N_{i}$
decrease exponentially as $\exp (\alpha \tau)$ and one increases
as $\exp (-\alpha \tau)$ for some $\alpha>0$ if $(\Sp,\Sm)$ is
in that neighbourhood. The solution must thus leave the neighbourhood
as we go backward in time
since otherwise Lemma \ref{lemma:B9N} or Lemma \ref{lemma:B8N} would
yield a contradiction. If the solution evaluated at $\tau_{k}$ converges
to the non-flat limit point on the Kasner circle, we can thus
construct a sequence $s_{k}$ such that the $\Sp\Sm$-variables of
the solution evaluated at it converges to a point on the boundary
of $B_{\epsilon}(\sigma_{+},\sigma_{-})$.
We can also assume that the $N_{i}$-variables of the solution,
evaluated at $s_{k}$, converge. The limit point will have at most one 
$N_{i}$-variable non-zero by the construction. If all the
$N_{i}$-variables are zero we have obtained a second $\alpha$-limit
point on the Kasner circle. If one is non-zero, we may apply the flow
to it to obtain two $\alpha$-limit points on the Kasner circle, 
cf. the proof of Theorem \ref{theorem:rendall}. Since the normalized
Kretschmann scalar is non-zero on a non-flat limit point on the Kasner
circle, see (\ref{eq:EH}), the last statement follows. $\Box$

\section{Existence of limit points}
In order to prove that an $\alpha$-limit point always exists we need
a few lemmas. Let $\|N\|$ denote the Euclidean norm of 
$N=(\No,\Nt,\Nth)$.

\begin{lemma}\label{lemma:nale}
Consider a Bianchi VIII or IX solution. If $\|N\|\rightarrow \infty$ 
as $\tau\rightarrow -\infty$ we can, by applying the symmetries to the
solution, assume that 
$\Nt,\Nth\rightarrow \infty$ and $\No,\No(\Nt+\Nth)\rightarrow 0$ 
as $\tau\rightarrow -\infty$.
\end{lemma}
\textit{Proof}. We have
\begin{equation}\label{eq:n123}
|\No\Nt\Nth|\leq C\ \ \forall\tau\leq 0
\end{equation}
where $C\geq 1$. Let $\tau_{0}\leq 0$ be such that
\begin{equation}\label{eq:ineq1}
\No^{2}+\Nt^{2}+\Nth^{2}\geq 300C^{2/3}\ \ \forall\tau\leq \tau_{0}.
\end{equation}
If for $\tau\leq \tau_{0}$ $|\No|=|\Nt|\leq|\Nth|$
we get $\Nth\geq 10C^{1/3}$ and $|\No|=|\Nt|\geq C^{1/3}$ by 
inequality (\ref{eq:ineq1}) and Lemma \ref{lemma:B9N} or 
\ref{lemma:B8N}. Since this 
is not reconcilable with inequality (\ref{eq:n123}) we conclude that 
one $N_{i}$, say $\No$, must satisfy $|\No|<\Nt$ and 
$|\No|<\Nth$ $\forall \tau\leq \tau_{0}$. For a Bianchi VIII solution
$\No<0$. But then $\Nt^{2}+\Nth^{2}\leq 101\Nt^{2}$ by Lemma
\ref{lemma:B9N} or \ref{lemma:B8N} and similarly for $\Nth$. 
Consequently $\Nt,\Nth \rightarrow \infty$ as $\tau\rightarrow
-\infty$. The rest of the conclusions of the lemma follow by
inequality (\ref{eq:n123}). $\Box$

\begin{lemma}\label{lemma:spinc}
Consider a Bianchi IX solution. If $-1<\Sp(\tau_{1})<1/3$ and 
$9\No<\Nt+\Nth$ in $[\tau_{2},\tau_{1}]$
then $\Sp(\tau_{1})\leq \Sp(\tau)$ for $\tau\in[\tau_{2},\tau_{1}]$.
\end{lemma}
\textit{Proof}. Using the constraint (\ref{eq:constraint}) we have
\begin{equation}\label{eq:spder}
\Sp'=(q-2)(\Sp+1)+\frac{9}{2}(\No^{2}-\No(\Nt+\Nth)).
\end{equation}
The constraint also yields $q-2\leq 3\No(\Nt+\Nth)$
so that, if $-1<\Sp<1/3$,
\[
\Sp'<\frac{1}{2}(9\No-\Nt-\Nth)\No.
\]
The lemma follows.$\Box$

\begin{lemma}\label{lemma:spinc2}
Consider a Bianchi VIII solution to
(\ref{eq:whsu})-(\ref{eq:constraint}) and let $I=[t,s]$, where
$s\leq0$ and $0$ belongs to the existence interval. There is an
$\epsilon_{0}>0$ with the following property: for all
$\epsilon_{0}>\epsilon>0$ there is an  $A>0$ such that if
$\Nt(\tau),\Nth(\tau)\geq A$ for all $\tau\in I$, then 
$-1+2\epsilon\leq\Sp(s)$ implies 
$-1+\epsilon\leq \Sp(\tau)$ for all $\tau\in I$.
$A$ only depends on $\epsilon$ and the solution.
\end{lemma}
\textit{Proof}. Let $\epsilon>0$. We have $|\No\Nt\Nth|\leq C$
in $I$ where $C$ only depends on the solution. Consider
\begin{equation}\label{eq:spder2}
\Sp'=(q-2)(\Sp+1)+\frac{9}{2}(\No^{2}-\No(\Nth+\Nth)).
\end{equation}
Observe that the last term is the only one pushing $\Sp$ in the
negative direction, since $q\leq 2$ and $\No<0$ for Bianchi VIII.
In order for $\Sp$ to attain $-1+\epsilon$ in $I$ it must go between
$-1+2\epsilon$ and $-1+\epsilon$ in $I$. Due to 
(\ref{eq:whsu}) the last term of the right hand side of
(\ref{eq:spder2}) will then behave as $\exp (2\tau-2\tau_{1})\eta$, 
if $\epsilon$ is small enough, where $\tau_{1}$ corresponds to 
$-1+2\epsilon$ and $\eta$ may be chosen arbitrarily small by choosing 
$A$ large enough. In consequence it is impossible for $\Sp$ to attain 
$-1+\epsilon$ if we choose $A$ big enough depending only on $\epsilon$ 
and the solution. $\Box$

\begin{prop}\label{prop:ale}
A Bianchi IX solution has an $\alpha$-limit point.
\end{prop}
\textit{Proof}. Assume $\|N\|\rightarrow \infty$. By Lemma 
\ref{lemma:nale} we may assume $\No,\No(\Nt+\Nth)\rightarrow 0$ and
$\Nt,\Nth\rightarrow \infty$ as $\tau\rightarrow -\infty$. Since 
$\Nt\Nth\rightarrow \infty$ we conclude that
\begin{equation}\label{eq:intest1}
\int_{0}^{\tau}(\Sp^{2}+\Sm^{2}+\Sp)d s\rightarrow \infty
\end{equation}
but then
\begin{equation}\label{eq:intest2}
\int_{\tau}^{0}\Sp d s\rightarrow -\infty.
\end{equation}
Since $9\No<\Nt+\Nth$ for $\tau\leq T$ Lemma \ref{lemma:spinc}
and (\ref{eq:intest2}) yield the conclusion that $\Sp<0\ \forall 
\tau\leq T$. 

Assume there is a $\tau_{2}\leq T$ such that $-1<\Sp(\tau_{2})$. 
Then $\Sp(\tau_{2})\leq \Sp(\tau)\ \forall \tau\leq \tau_{2}$ by 
Lemma \ref{lemma:spinc}. Let 
$\mathcal{A}=\{\tau:\ \Sp^{2}+\Sm^{2}<-\Sp\}$
and $\mathcal{A}_{\tau}=\mathcal{A}\cap[\tau,\tau_{2}]$. Since
$\Sp'<0\ \forall\tau\leq \tau_{2}$, cf. the proof of Lemma
\ref{lemma:spinc}, we have by (\ref{eq:spder})
\begin{eqnarray*}
\Sp(\tau_{2})-\Sp(\tau)\leq
\int_{\mathcal{A}_{\tau}}\Sp'd s\leq\
(1+\Sp(\tau_{2}))\int_{\mathcal{A}_{\tau}}(q-2)d s\\
\leq 2(1+\Sp(\tau_{2}))\int_{\mathcal{A}_{\tau}}(\Sp^{2}+\Sm^{2}+
\Sp)d s.
\end{eqnarray*}
By (\ref{eq:intest1}) the expression on the right tends to 
$-\infty$ as $\tau\rightarrow -\infty$ so that $\Sp(\tau)
\rightarrow \infty$ which is a contradiction.

Assume there is an $S$ such that $\Sp(\tau)\leq -1\ \forall \tau\leq
S$. Then the constraint yields $(\Sp,\Sm)\rightarrow (-1,0)$ and
Proposition \ref{prop:limchar} yields $\Sm=0$, so that all the
$N_{i}$ must decrease for $\tau\leq S$ as we go backward in time by 
(\ref{eq:whsu}) and (\ref{eq:whsudef}). We have a contradiction to 
our assumption. 

In other words there is a sequence $\tau_{k}$ such that
$N_{i}(\tau_{k})$ is bounded. Since $q(\tau_{k})$ is contained in a 
compact set by Lemma \ref{lemma:B9N} we may extract a convergent 
subsequence and the proposition follows. $\Box$

\begin{cor}\label{cor:shearcon}
Consider a Bianchi IX solution. For all $\epsilon>0$ there is a $T$
such that $\tau\leq T$ implies 
\[
q(\tau)\leq 2+\epsilon.
\]
\end{cor}
\textit{Remark}. For a Taub-NUT solution there is a $T$ such that
$q(\tau)>2$ for all $\tau\leq T$.

\textit{Proof}. Since there is an $\alpha$-limit point, Lemma
\ref{lemma:onz} proves that there is a sequence $\tau_{k}$ such
that $(\No\Nt\Nth)(\tau_{k})\rightarrow 0$ but since $\No\Nt\Nth$
is monotone we may conclude that the product converges to zero.
Let $10^{-3}>\eta>0$. Let $T$ be such that $(\No\Nt\Nth)(\tau)\leq
\eta^{3}$ for all $\tau\leq T$ and assume for a $\tau\leq T$ that
$\Nth\geq\Nt\geq\No$. We get $\No(\tau)\leq\eta$. If, in $\tau$, 
$\Nth\geq 2$ Lemma \ref{lemma:B9N} yields   
$\No(\Nt+\Nth)\leq \eta$. If $\Nth\leq 2$ in $\tau$ we get
$\No(\Nt+\Nth)\leq 4\eta$. Thus we have 
\[
q\leq 2+12\eta
\]
and the statement follows by the constraint
(\ref{eq:constraint}). $\Box$

\begin{prop}
A Bianchi VIII solution has an $\alpha$-limit point.
\end{prop}
\textit{Proof}. Assume $\|N\|\rightarrow \infty$ as $\tau\rightarrow
-\infty$. The consequences given in Lemma \ref{lemma:nale} will be
used freely below. As $\Nt\Nth\rightarrow \infty$ we get
\begin{equation}\label{eq:intest3}
\int_{\tau}^{0}(\Sp^{2}+\Sm^{2}+\Sp)d s\rightarrow -\infty
\end{equation}
by inspecting (\ref{eq:whsu}). If $\Sp\rightarrow -1$ Proposition 
\ref{prop:limchar} yields $\Nt=\Nth$ and $\Sm=0$. By the constraint
and Lemma \ref{lemma:limestimate}, $1+\Sp$ decays exponentially so 
that all the $N_{i}$ converge to finite values, contradicting our 
assumption. If not there is an $\epsilon>0$ and  by Lemma 
\ref{lemma:spinc2} a $T$ such that $\tau\leq T$ implies 
$-1+\epsilon/2\leq \Sp(\tau)$. We get, sooner or later,
\begin{eqnarray*}
\Sp'=(q-2)(\Sp+1)+\frac{9}{2}(\No^{2}-\No(\Nt+\Nth))
\leq (\Sp^{2}+\Sm^{2}+\Sp-\Sp-1)\epsilon\\
+\frac{9}{2}(\No^{2}-\No(\Nt+\Nth))
\leq(\Sp^{2}+\Sm^{2}+\Sp)\epsilon
\end{eqnarray*}
so that
\[
\Sp(\tau_{2})-\Sp(\tau)\leq\epsilon
\int_{\tau}^{\tau_{2}}(\Sp^{2}+\Sm^{2}+\Sp)d s\rightarrow -\infty
\]
by (\ref{eq:intest3}). But that is not possible. The existence of
an $\alpha$-limit point follows as in the previous proposition.$\Box$

\begin{thm}\label{thm:alp}
If a Bianchi VIII or IX solution to
(\ref{eq:whsu})-(\ref{eq:constraint}) is not 
contained in the invariant set $\Sm=0,\ \Nt=\Nth$ or one of the
sets found by applying the symmetries to it, the solution has an 
$\alpha$-limit point $(n_{1},n_{2},n_{3},s_{+},s_{-})$ for which 
$(s_{+},s_{-})$ is not special.
\end{thm}
\textit{Proof}. Consider a solution not contained in the invariant
sets mentioned and let $(n_{1},n_{2},n_{3},s_{+},s_{-})$ be an 
$\alpha$-limit point. If $(s_{+},s_{-})$ is not special we are done, 
so assume $(s_{+},s_{-})=(-1,0)$. Let $\tau_{k}$ be a sequence such
that the solution evaluated at $\tau_{k}$ converges to the given
$\alpha$-limit point. Since $(\Sp,\Sm)$ does not converge to the 
special point there is an $0<\epsilon<10^{-3}$ such that for 
$k$ large enough we may for each $\tau_{k}$ find the first time,
call it $s_{k}\leq\tau_{k}$, that $(\Sp(s_{k}),\Sm(s_{k}))\in 
\partial B_{\epsilon}(-1,0)$. If there is a subsequence of $s_{k}$
such that $N_{i}(s_{k})$ is bounded we are done so assume not.

The goal is to go backward in time starting at $s_{k}$ in order to
produce a sequence with the desired properties.
Since $q$ is bounded as we go backward in time we must have 
$\tau_{k}-s_{k}\rightarrow \infty$ so that $\No(s_{k})
\rightarrow 0$ since $\No'=(q-4\Sp)\No$ and $q-4\Sp$ is roughly 
$6$ in $B_{\epsilon}(-1,0)$. We get $\Nth(s_{k}),\Nt(s_{k})
\rightarrow \infty$ and $\No(\Nt+\Nth)\rightarrow 0$ in $s_{k}$.

\textit{Bianchi IX}. There is an $\eta>0$ such that 
$-1+\eta\leq\Sp(s_{k})\leq 0$
by the constraint if $k$ great enough, and then we also have
$9\No(s_{k})<\Nt(s_{k})+\Nth(s_{k})$.
If the latter inequality holds $\forall \tau\leq s_{k}$ we get a
contradiction to the existence of our limit point by Lemma
\ref{lemma:spinc}. Let $v_{k}\leq s_{k}$ be the first time
$9\No=\Nt+\Nth$ or one of $\Nt,\ \Nth$ becomes equal to $\No$.
Since $\No\Nt\Nth\rightarrow 0$, cf. the proof of Corollary 
\ref{cor:shearcon}, we may assume 
\begin{equation}\label{eq:ncon}
\No\Nt\Nth\leq 1.
\end{equation}
If $\No(v_{k})=\Nt(v_{k})\leq \Nth(v_{k})$ then
$\No(v_{k})=\Nt(v_{k})\leq 1$ by (\ref{eq:ncon}) and $\Nth(v_{k})\leq
10$ by Lemma \ref{lemma:B9N} and similarly if $\Nt$ and $\Nth$
change roles. If $\No(v_{k})<N_{i}(v_{k})$ $i=2,3$ and
$9\No(v_{k})=\Nt(v_{k})+\Nth(v_{k})$, then $\No(v_{k})\leq 1$
and $N_{i}(v_{k})\leq 9$, $i=2,3$. Regardless, we have 
$N_{i}(v_{k})\leq 10$. For $k$ great enough there
is a $t_{k}$ and a $u_{k}$ with $v_{k}\leq u_{k}\leq t_{k}\leq s_{k}$
such that $\Nth(t_{k})=10^{20}$, $\Nth(u_{k})=10^{10}$ and
\begin{equation}\label{eq:n3con}
10^{10}\leq\Nth(\tau)\leq 10^{20}\ \ \forall \tau\in[u_{k},t_{k}].
\end{equation}
In consequence
\begin{equation}\label{eq:n2con}
10^{9}\leq\Nt(\tau)\leq 10^{21}\ \ \forall\tau\in [u_{k},t_{k}]
\end{equation}
by Lemma \ref{lemma:B9N}. The inequalities above are used to prove
that $t_{k}-u_{k}\geq 1$ from which we deduce the existence of an
$r_{k}\in[u_{k},t_{k}]$ such that $|\Sm(r_{k})|\leq 1/10$. That
will suffice to prove the lemma.

We get
\[
20\leq\ln\frac{\Nth(t_{k})}{\Nth(u_{k})}=
\int_{u_{k}}^{t_{k}}(q+2\Sp-2\sqrt{3}\Sm)d s.
\]
But $\No(\Nt+\Nth)\leq 2\cdot 10^{-9}$ in $[u_{k},t_{k}]$  by
(\ref{eq:ncon})-(\ref{eq:n2con}) so that
$q\leq 3$ in that interval by the constraint (\ref{eq:constraint}), 
and thus $t_{k}-u_{k}\geq 1$. We have $|\Nt/\Nth-1|\leq 2\cdot
10^{-10}$ in $[u_{k},t_{k}]$ by the constraint and the above.
If $|\Sm|\geq 1/10$ in $[u_{k},t_{k}]$ we get
\[ 
|\int_{u_{k}}^{t_{k}}4\sqrt{3}\Sm d\tau|\geq \frac{3}{5}.
\]
But
\[
\frac{\Nt(t_{k})}{\Nth(t_{k})}=\exp (\int_{u_{k}}^{t_{k}}4\sqrt{3}
\Sm d\tau) \frac{\Nt(u_{k})}{\Nth(u_{k})}
\]
and we have a contradiction. Consequently there is an
$r_{k}\in[u_{k},t_{k}]$ such that $-1+\eta\leq \Sp(r_{k})$,
$|\Sm(r_{k})|\leq 1/10$ and $N_{i}(r_{k})\leq 10^{21}$. The 
conclusion of the theorem follows. 

\textit{Bianchi VIII}. There is an $\eta>0$ such that 
$-1+2\eta\leq\Sp(s_{k})\leq 0$. Let $A=A(\eta)\geq 1$ be as in Lemma 
\ref{lemma:spinc2} and $10A\leq\Nth(\tau)\ \
\forall\tau\in[v_{k},s_{k}]$ and $\Nth(v_{k})=10A$, observe that by
the existence of the limit point the $N_{i}$ must become small.
Then we may argue as in the Bianchi IX case.
$\Box$

\section{Conclusions}
\begin{thm}\label{thm:blowup}
Consider a Bianchi VIII or IX solution of
(\ref{eq:whsu})-(\ref{eq:constraint}) which is not of NUT or Taub-NUT
type respectively. Then there is a sequence $\tau_{k}\rightarrow
-\infty$ such that $\tilde{\kappa}(\tau_{k})\rightarrow c\neq 0$.
\end{thm}
\textit{Proof}. The Taub-NUT and the NUT solutions correspond to 
$\Sm=0$ and $\Nt=\Nth$ or one of the sets obtained by applying
the symmetries. By Theorem \ref{thm:alp} and 
Lemma \ref{lemma:knz} the theorem follows. $\Box$

\begin{thm}
For a Bianchi VIII or IX solution to
(\ref{eq:whsu})-(\ref{eq:constraint}) which is not a NUT or Taub-NUT
solution $(\Sp,\Sm)$ cannot converge to a special point on the 
Kasner circle and the $\alpha$-limit set contains at least two 
distinct points of type I, at least one of which is non-flat.
\end{thm}
\textit{Proof}. The first statement is Proposition \ref{prop:limchar}
and the last statement follows from Theorem \ref{thm:alp} and
Lemma  \ref{lemma:knz}. $\Box$

Observe that this theorem says that the solution does not converge;
the shear variables will oscillate indefinitely.

\textit{Proof of Theorem \ref{thm:main}}. Let $(M,g)$ be the globally 
hyperbolic Lorentz manifold obtained in Lemma \ref{lemma:development}. 
Assume there is a connected Lorentz manifold 
$(\hat{M},\hat{g})$ of the same dimension and a map 
$i:M\rightarrow \hat{M}$ which is 
an isometry onto its image  with $i(M)\neq \hat{M}$. Then there
is a $p\in \hat{M}-i(M)$ and a timelike geodesic
$\gamma:[a,b]\rightarrow \hat{M}$ such that $\gamma([a,b))\subseteq
i(M)$ and $\gamma(b)=p$. Since $\gamma |_{[a,b)}$ can be considered
to be a future or past inextendible timelike geodesic in $M$ either it
has infinite length or the Kretschmann scalar blows up along it,
combining Lemma \ref{lemma:development}, \ref{lemma:emwh} and
Theorem \ref{thm:blowup}. Both possibilities lead to a contradiction. 
Observe that this proves that $(M,g)$ is the maximal globally 
hyperbolic development.  That the Kretschmann scalar is unbounded
in the incomplete directions of inextendible causal geodesics also
follows from Lemma \ref{lemma:development}, \ref{lemma:emwh} and
Theorem \ref{thm:blowup}. $\Box$

\appendix

\section*{Appendix}

\setcounter{section}{1}

We here prove the technical results of Lemma \ref{lemma:development}
and \ref{lemma:emwh}.

\textit{Proof of Lemma \ref{lemma:development}}. We begin by proving
that we obtain a solution to Einstein's vacuum equation with the 
correct initial conditions. Let $e_{i}'$, $i=1,2,3$ 
be a left invariant orthonormal basis. We can assume the corresponding 
$n'$ to be of one of the
forms given in table \ref{table:bianchiA} by Lemma \ref{lemma:liealg}. 
The content of (\ref{eq:con2}) is that $k_{ij}=k(e_{i}',e_{j}')$ and 
$n'$ are to commute. We may thus also assume $k_{ij}$ to 
be diagonal without changing the earlier conditions of the
construction. If we let $n(t_{0})=n'$, $\theta(t_{0})=
\mathrm{tr}_{g}k$ and $\sigma_{ij}(t_{0})=k_{ij}-\theta\delta_{ij}/3$ 
then (\ref{eq:con1}) is the same as (\ref{eq:constraint1}). Let $n$,
$\sigma$ and $\theta$ satisfy (\ref{eq:dndt}), (\ref{eq:dsdt}) and
(\ref{eq:raychaudhuri}) with initial values as specified above. 
Since (\ref{eq:constraint1}) is satisfied at $t_{0}$ it is satisfied 
for all times. For reasons given in connection with
(\ref{eq:constraint1}) $n$ and $\sigma$ will 
remain diagonal so that (\ref{eq:commute2}) will always hold. 

Let $M=I\times G$ where $I$ is the maximal existence interval for
solutions to (\ref{eq:dndt})-(\ref{eq:constraint1}). We construct a
basis $e_{\alpha}$, define a metric by demanding that the basis
be orthonormal and show that the corresponding $\tilde{n}$, 
$\tilde{\sigma}$ and $\tilde{\theta}$ coincide with $n$, $\sigma$ 
and $\theta$. We will thereby have constructed a Lorentz manifold 
satisfying Einstein's vacuum equations with the correct initial 
conditions.

Let $n_{i}$ and $\sigma_{i}$ denote the diagonal elements of $n$ and
$\sigma$ respectively. Let $f_{i}(t_{0})=1$ and
$\dot{f_{i}}/f_{i}=2\sigma_{i}-\theta/3$.
Let $a_{1}=(f_{2}f_{3})^{1/2}$, 
$a_{2}=(f_{1}f_{3})^{1/2}$, $a_{3}=(f_{1}f_{2})^{1/2}$ and
define $e_{i}=a_{i}e_{i}'$. Then $\tilde{n}$ associated to $e_{i}$ 
equals $n$. We complete the basis by letting $e_{0}=\partial_{t}$. 
Define a metric $<\cdot,\cdot>$ on $M$ by demanding $e_{\alpha}$ to be 
orthonormal with $<e_{0},e_{0}>=-1$ and $<e_{i},e_{i}>=1$, $i=1,2,3$
and let $\nabla$ be the associated Levi-Civita
connection. Compute $<\nabla_{e_{0}}e_{i},e_{j}>=0$.
If $\tilde{\theta}(X,Y)=<\nabla_{X}e_{0},Y>$ and
$\tilde{\theta}_{\mu\nu}=\tilde{\theta}(e_{\mu},e_{\nu})$, then 
$\tilde{\theta}_{00}=\tilde{\theta}_{i0}=
\tilde{\theta}_{0i}=0$. Furthermore,
\[
\frac{1}{a_{j}}e_{0}(a_{j})\delta_{ij}=-\tilde{\theta}_{ij}
\]
(no summation over $j$) so that $\tilde{\theta}_{ij}$ is diagonal and
$\mathrm{tr}\tilde{\theta}=\theta$. Finally,
\[
-\tilde{\sigma}_{ii}=-\tilde{\theta}_{ii}+\frac{1}{3}\theta=-\sigma_{i}.
\]
The constructed Lorentz manifold thus satisfies Einstein's 
vacuum equations. Next we prove that each $M_{v}=\{ v\}\times G$ is
a Cauchy surface. The metric is given by
\[
-d t^{2}+\sum_{i=1}^{3}a_{i}^{-2}(t)\xi^{i}\otimes \xi^{i}
\]
where $\xi^{i}$ are the duals of $e_{i}'$. A causal curve cannot
intersect $M_{v}$ twice since the $t$-component of such a curve must
be strictly monotone. Assume that $\gamma:(s_{-},s_{+})\rightarrow M$ 
is an inextendible causal curve that never intersects $M_{v}$. Let 
$t:M\rightarrow I$ be defined by $t[(s,h)]=s$. Let
$s_{0}\in(s_{-},s_{+})$ and assume that  
$t(\gamma(s_{0}))=t_{1}<v$ and that $<\gamma',\partial_{t}><0$ where 
it is defined. Thus $t(\gamma(s))$ increases with $s$ and 
$t(\gamma([s_{0},s_{+})))\subseteq [t_{1},v]$. Since we have
uniform bounds on $a_{i}$ from below and above on $[t_{1},v]$ and 
the curve is causal we get 
\[
(\sum_{i=1}^{3}\xi^{i}(\gamma')^{2})^{1/2}\leq -C<\gamma',e_{0}>
\]
on that interval, with $C>0$. Since 
\[
\int_{s_{0}}^{s_{+}}-<\gamma',e_{0}>d s=\int_{s_{0}}^{s_{+}}
\frac{d t\circ \gamma}{d s}d s\leq v-t_{1}
\]
the curve $\gamma|_{[s_{0},s_{+})}$, projected to $G$, will have finite 
length in the
metric $\rho$ on $G$ defined by making $e_{i}'$ an orthonormal basis.
Since $\rho$ is a left invariant metric on a Lie group it is complete 
 and sets closed and bounded in the
corresponding topological metric must be compact. Adding the 
above observations, we conclude that $\gamma([s_{0},s_{+}))$ is 
contained in a compact set. For each  sequence  
$s_{k}\rightarrow s_{+}-$ there is thus a subsequence $s_{n_{k}}$ 
such that $\gamma(s_{n_{k}})$ converges. Since $t(\gamma(s))$ is
monotone it converges. We cannot have two limit points
since that would contradict the causality of $\gamma$. Thus 
$\gamma$ must converge as $s\rightarrow
s_{+}$ so that it is extendible. By this and similar arguments
covering the other cases, we conclude that $M_{v}$
is a Cauchy surface for each $v\in (t_{-},t_{+})$. 

Next we prove the statements made in Lemma \ref{lemma:development}
concerning causal geodesic completeness. Let us first time orient 
the different manifolds.
Consider manifolds which are not of type IX. Then the constraint
(\ref{eq:constraint1}) yields the inequality $\sigma_{ij}\sigma^{ij}
\leq 2\theta^{2}/3$, cf. (\ref{eq:shearcon}). Then the Raychaudhuri
equation yields $|e_{0}(\theta)|\leq \theta^{2}$. 
Consequently, if $\theta$ is once zero it is always zero. By
considering equations (\ref{eq:dndt})-(\ref{eq:constraint1})
one concludes that $\theta=0$ is only possible if
$\sigma_{ij}=n_{ij}=0$ or if two of the diagonal elements of $n_{ij}$
are equal and constant, the third is zero and $\sigma_{ij}=0$. Time 
orient the manifolds which are not of type IX and which do not have 
$\theta=0$ by demanding that $\theta$ be positive.

Observe that for a manifold which is not of type IX, $\theta$
decreases in magnitude with time, so that it is bounded to the future.
By the constraint (\ref{eq:constraint1}), the same is true of
$\sigma_{ij}$. Using (\ref{eq:dndt}) we get control of $n_{ij}$
and conclude that the solution may not blow up in finite time.
The interval $I=(t_{-},t_{+})$ in Lemma \ref{lemma:development}
must thus have $t_{+}=\infty$. 

We now prove future causal geodesic completeness for manifolds which
are not of type IX. Let $\gamma:(s_{-},s_{+})\rightarrow M$ be a 
future directed inextendible causal geodesic. We prove that $s_{+}$
must be infinite. Let the function $t$
be defined as in the previous lemma. Since every $M_{v}$,
$v\in I$ is a Cauchy surface, $t(\gamma(s))$ must cover the interval
$I$ as $s$ runs through $(s_{-},s_{+})$. Furthermore, $t(\gamma(s))$
is monotone increasing. Define
\[
f_{\mu}(s)=<\gamma'(s),e_{\mu}|_{\gamma(s)}>.
\]
Let $s_{0}\in (s_{-},s_{+})$ and compute
\begin{equation}\label{eq:intf}
\int_{s_{0}}^{s}-f_{0}(u)d u=t(\gamma(s))-t(\gamma(s_{0}))
\end{equation}
so that the right hand side goes to $\infty$ as $s\rightarrow s_{+}$.
If we can prove that $f_{0}$ may not become unbounded in a finite
$s$-interval, we are done. If $\theta\equiv 0$, then $(t_{-},t_{+})=
(-\infty,\infty)$ and $f_{0}$ is constant so that all causal geodesics
are future and past complete. We exclude this case from now on. Compute
\[
\frac{d f_{0}}{d s}=<\gamma'(s),\nabla_{\gamma'(s)}e_{0}>=
\sum_{k=1}^{3}\theta_{k}f_{k}^{2}
\]
where $\theta_{k}$ are the diagonal elements of $\theta_{ij}$.
Consider functions of $t$ as functions of $s$ by evaluating
them at $t(\gamma(s))$. Compute,
using Raychaudhuri's equation (\ref{eq:raychaudhuri}),
\[
\frac{d}{d s}(f_{0}\theta)=
\frac{1}{3}\theta^{2}\sum_{k=1}^{3}f_{k}^{2}+\sum_{k=1}^{3}\theta
\sigma_{k}f_{k}^{2}+f_{0}^{2}\sum_{k=1}^{3}\sigma_{k}^{2}+
\frac{1}{3}\theta^{2}f_{0}^{2}
\]
where $\sigma_{k}$ are the diagonal elements of $\sigma_{ij}$.
Estimate
\[
|\sum_{k=1}^{3}\sigma_{k}f_{k}^{2}|\leq \left(\frac{2}{3}\right)^{1/2}
\left(\sum_{k=1}^{3}\sigma_{k}^{2}\right)^{1/2}\sum_{k=1}^{3}f_{k}^{2}
\]
using the tracelessness of $\sigma_{ij}$. By making a division into
the three cases $\sum_{k=1}^{3}\sigma_{k}^{2}\leq \theta^{2}/3$,
$\theta^{2}/3\leq \sum_{k=1}^{3}\sigma_{k}^{2}\leq 2\theta^{2}/3$
and $2\theta^{2}/3\leq \sum_{k=1}^{3}\sigma_{k}^{2}$ and using the
causality of $\gamma$ we deduce
\begin{equation}\label{eq:geoest}
\frac{d}{d s}(f_{0}\theta)\geq \frac{2-\sqrt{2}}{3}\theta^{2}
f_{0}^{2}.
\end{equation}
Observe that this estimate holds for a Bianchi IX solution as well.

Since $f_{0}\theta$ is negative on $[s_{0},s_{+})$, its absolute
value is thus bounded on that interval. If $s_{+}$ were finite,
$\theta$ would be bounded from below by a positive constant on 
$[s_{0},s_{+})$, since
\[
|\frac{d \theta}{d s}|\leq -f_{0}\theta^{2}\leq C\theta
\]
on that interval for some $C>0$. Then (\ref{eq:geoest}) would
imply the boundedness of $f_{0}$ on $[s_{0},s_{+})$. That
would contradict (\ref{eq:intf}). 

Excluding Bianchi IX for the moment, the inequality (\ref{eq:geoest}) 
also proves that all causal geodesics are past incomplete since
it implies that $f_{0}\theta$ must blow up after a finite $s$-time
going into the past. 

Consider a Bianchi IX manifold. By a theorem by Lin and
Wald (1989a), the trace of the second fundamental form will be 
zero for some $t_{0}\in I=(t_{-},t_{+})$ (in their paper they demand
that $G$ have topology $S^{3}$, but this is not necessary for their
argument). However, on $I_{-}=(t_{-},t_{0})$ and on 
$I_{+}=(t_{0},t_{+})$ 
$\theta\neq 0$ since it is only zero once (this can be
seen by inspecting equations (\ref{eq:dndt})-(\ref{eq:constraint1})). 
A future directed inextendible causal geodesic
$\gamma$ in a Bianchi IX manifold must consequently have points 
$s_{1},s_{2}$
with $\theta(t(\gamma(s)))$ positive (negative) for $s\leq s_{1}$
($s\geq s_{2}$). The inequality (\ref{eq:geoest}) now
implies that the geodesic cannot be defined more than a finite
interval before $s_{1}$ nor more than a finite interval after $s_{2}$. 
We have future and past causal geodesic incompleteness. We also conclude
that $t_{-}>-\infty$ for all manifolds that do not satisfy 
$\theta\equiv 0$
and $t_{+}<\infty$ for Bianchi IX manifolds since the curve defined
by $\gamma(s)=(s,e)$ is a geodesic.

Finally, we prove the last statement of Lemma \ref{lemma:development}.
Assume the Kretschmann scalar is unbounded as $t\rightarrow t_{-}$ and
let $\gamma$ be a past inextendible causal geodesic. Since each
$M_{v}$ is a Cauchy surface, $\gamma$ must pass through each of them
and thus the Kretschmann scalar must be unbounded along it. $\Box$

\textit{Proof of Lemma \ref{lemma:emwh}}. Consider a Bianchi IX
solution of (\ref{eq:dndt})-(\ref{eq:constraint1}). As observed in
the proof of Lemma \ref{lemma:development} the existence interval 
$I=(t_{-},t_{+})$ can be divided into $I_{-}=(t_{-},t_{0})$,
$I_{+}=(t_{0},t_{+})$ and $t_{0}$, where $t_{0}$ is the only zero
of $\theta$ in $I$. We now relate the different time coordinates on
$I_{-}$.

According to equation (\ref{eq:dtdtau}) $\tau$ has to satisfy
$d t/d \tau=3/\theta$. Define
$\tau(t)=\int_{t_{1}}^{t}\theta(s)/3d s$,
where $t_{1}\in I_{-}$. Then $\tau:I_{-}\rightarrow \tau(I_{-})$ is a 
diffeomorphism and strictly monotone on $I_{-}$. By equation 
(\ref{eq:raychaudhuri}) $\theta$ decreases so that it will be positive
in $I_{-}$ and $\tau$ will increase with $t$.

Since $\theta$ is continuous beyond $t_{0}$ it is clear that 
$\tau(t)\rightarrow \tau_{0} \in\mathbb{R}$ as $t\rightarrow t_{0}$.
To prove that $t\rightarrow t_{-}$ corresponds to $\tau\rightarrow
-\infty$ we make the following observation. One of the expressions 
$\theta$ and $d \theta/d t$ is unbounded on $(t_{-},t_{1}]$,
since if both were bounded the same would be true of $\sigma_{ij}$
and $n_{ij}$ by (\ref{eq:raychaudhuri}) and (\ref{eq:dndt})
respectively. Then we would be able to extend the solution beyond
$t_{-}$, contradicting the fact that $I$ is the maximal existence 
interval (observe that $t_{-}>-\infty$ by the proof of Lemma
\ref{lemma:development}). If $\tau$ were bounded from below on 
$I_{-}$, then $\theta$
and $\theta'$ would be bounded on $\tau((t_{-},t_{1}])$ by Lemma
\ref{lemma:existence}, and thus $\theta$ and $d \theta/d t$
would be bounded on $(t_{-},t_{1}]$. Thus $t\rightarrow t_{-}$
corresponds to $\tau\rightarrow -\infty$. Since $\theta\rightarrow
\infty$ as $\tau\rightarrow -\infty$ by (\ref{eq:raychaudhuri2}),
$\kappa(\tau_{k})=\tilde{\kappa}(\tau_{k})\theta^{4}(\tau_{k})$ is
unbounded. Consequently
\[
\limsup_{t\rightarrow t_{-}}|R_{\alpha\beta\gamma\delta}
R^{\alpha\beta\gamma\delta}|=\infty
\]
for all non-Taub-NUT initial data given the assumptions of the Lemma. 
Similar arguments yield the same conclusion for $t\rightarrow
t_{+}$ and for $t\rightarrow t_{-}$ given non-NUT Bianchi VIII initial
data. $\Box$

\section*{Acknowledgments}
The author would like to express his gratitude to his advisor, Lars
Andersson, for suggesting the problem and helpful discussions, and
to Alan Rendall for reading the argument and suggesting improvements.

\end{document}